# How universal is the Lindemann criteria in melting of Lennard-Jones polydisperse solids?

Sarmistha Sarkar, Chandramohan Jana and Biman Bagchi[*]

*Solid State and Structural chemistry Unit, Indian Institute of Science, Bangalore 560012, India*

*Email: profbiman@gmail.com, bbagchi@sscu.iisc.ernet.in*

## Abstract

**It is commonly believed that melting occurs when mean square displacement (MSD) of a particle of crystalline solid exceeds a threshold value. This is known as the Lindemann criterion, first introduced in the year of 1910 by Lindemann. However, Chakravarty *et al.* demonstrated that this common wisdom is inadequate because the MSD at melting can be temperature dependent when pressure is also allowed to vary along the coexistence line of the phase diagram [Chakravarty C, Debenedetti P G and Stillinger F H 2007 *J. Chem. Phys.* 126 204508]. We show here by extensive molecular dynamics simulation of both two and three dimensional polydisperse Lennard-Jones solids that particles on the small and large limits of size distribution exhibit substantially different Lindemann ratio at melting. Despite all the dispersion in MSD, melting is found to be first order in both the dimensions at 5-10% dispersity in size.**

# Introduction

Lindemann criterion holds a special place in the study of melting transition and is often believed to provide a universal criterion for the solid-liquid transition. The Lindemann criterion is the only accessible predictive tool in understanding the solid-liquid coexistence conditions for several transition metals and geologically significant minerals possessing high melting temperatures [1, 2].

The Lindemann criterion states that during the melting of solid, the average amplitude of thermal vibrations increases with increase in temperature and melting occurs if the amplitude of vibration becomes large enough for displacements of atoms compared to their equilibrium lattice sites which are in the range of one-half of the interatomic distance [3]. Later the Lindemann parameter was modified by Gilvarry [4] by considering the root-mean-square amplitude of thermal vibrations. According to Gilvarry, the melting process is initiated when the fraction of root-mean-square amplitude and interatomic distance reaches a critical value. The critical value of Lindemann parameter is basically taken as 0.1, but the value may vary ranging from 0.05 to 0.20 depending on the factors such as nature of interparticle interactions, magnitude of quantum effects and crystal structure [5-7].

In a recent study [8], in order to investigate the effect of polydispersity on the solid-liquid transition of the Lennard-Jones system [9], we have analyzed the model system using numerical simulations in three distinct ways, namely, by applying the empirical Lindemann criterion of melting, the inherent structure analysis [10-12] and the empirical Hansen-Verlet criterion of freezing [13, 14]. The result has been shown to be consistent with one another, both qualitatively and quantitatively, in predicting the existence of terminal polydispersity, beyond which freezing is not possible [9]. The Lindemann ratio, obtained from RMSD, has been shown to be intensely dependent on temperature [8].

Unlike a one component system, polydisperse liquid or solid possesses distinguishable molecules having different sizes. Therefore the fascinating characteristics related to the contribution of particles on the small and large limits of size distribution of polydisperse solids (will be termed henceforth as small and big particles respectively) during the melting transition is a matter of great interest [15] in the field of polydispersity.

In addition to its fundamental interest, a polydisperse system is of considerable industrial and technological importance. Over the last few decades polydisperse fluids have drawn considerable attention by the scientific community. As a result a numerous number of numerical, analytical as well as experimental studies have been reported on polydisperse fluids [16-32]. However, there are several issues yet to be explored. One of the significant ways to explore the range of stability of a crystalline solid is to study the effect of Lindemann criterion during the melting of polydisperse systems capable of providing comprehensive understanding of solid-liquid coexistence during the melting transition.

Chakravarty, Debenedetti and Stillinger [1] considered the Lindemann ratio within the context of inherent structure in order to understand the solid-liquid phase transition in a Lennard-Jones-type system. A valuable synopsis for the properties of coexisting fluid and solid phases in the case of soft sphere was presented by Kofke and co-workers [6]. Several semi-empirical 'melting rules' were examined by Kofke *et al.* in the light of the results. Moreover, Bolhuis and Kofke have revealed the existence of terminal polydispersity for the solid phase to be 5.7% and for the liquid phase to be 11.8% that is in good agreement with our earlier work [9].

In the present study we demonstrate that for a polydisperse solid consisting of particles with different sizes interacting with Lennard-Jones interaction potential, the Lindemann constant is found to be size dependent giving rise to different values for large and small sized particles. The breakdown of the universal Lindemann criterion is observed both in the cases of two and three dimensional polydisperse systems during melting transitions.

The organization of the rest of the paper is as follows. In Sec. 2, we illustrate the computational details and the model system. In Sec. 3, we organize the results in different subsections and elucidate precise discussion of the results. In Sec. 4, we present concluding remarks on the work.

1. Model and simulation details

In this study, molecular dynamics (MD) simulations of size dispersed Lennard-Jones (LJ) particles are carried out both in two-dimensional (2D) and three-dimensional (3D) systems. The simulations are performed under periodic boundary condition for total number of particles N = 2500 particles in case of 2D and N = 1372 particles in case of 3D system. The particles interact via the Lennard-Jones 12-6 potential,

$$U(r) = 4\varepsilon_{ij}\left[\left(\frac{\sigma_{ij}}{r}\right)^{12} - \left(\frac{\sigma_{ij}}{r}\right)^{6}\right]$$

where r is the distance between two particles and $i, j$ represent any two particles. The particle diameter is given by, $\sigma_{ij} = (\sigma_i + \sigma_j)/2$. In our calculations we have truncated the Lennard-Jones potential at a distance of $2.5\sigma_{ij}$ and potential is shifted to zero at $2.5\sigma_{ij}$. The potential well depth $\varepsilon_{ij}$ is considered as identical for all particle pairs and set to unity. The size dispersity is introduced by using the random sampling from the Gaussian distribution of particle diameter $\sigma$,

$$p(\sigma) = \frac{1}{\sqrt{2\pi d^2}} \, exp\left[-\frac{1}{2}\left(\frac{\sigma - \bar{\sigma}}{d}\right)^2\right]$$

where, $d$ is the standard deviation of the distribution and $\bar{\sigma}$ is the mean diameter. The polydispersity index is defined as $\delta = \frac{d}{\bar{\sigma}}$.

We have performed molecular dynamics (MD) simulations in constant temperature and pressure (NPT) ensemble. The simulations are performed both in the cases of 2D and 3D systems for the value of polydispersity index $\delta = 0.05$. The step-wise heating is performed for each 1000000 MD time steps and after each step, the temperature ($T^*$) is increased by 0.02 and heating is continued until the solid phase changes to liquid phase. During the heating process, we have calculated density at each temperature and in the whole process, we have kept a fixed pressure $P^* = 10$.

Additionally, in order to study the finite size effect we have performed simulations for two other systems having sizes N = 6400 and N = 5324 for 2D and 3D polydisperse systems respectively. However, there is no considerable difference observed with the increase in the system sizes. In the forthcoming section, we present the results related to the system size of N = 2500 particles for two dimensional and N=1372 particles for three dimensional polydisperse systems.

## 2. Result and discussion

### 2.1. Temperature induced solid-liquid melting transition in 2D and 3D polydisperse systems

#### 2.1.1. Temperature-density $(T - \rho)$ phase diagram

In the temperature induced solid-liquid melting transition, the density of solid gradually decreases with increasing temperature and after a certain temperature, the solid goes to the liquid phase. **Fig. 1** represents temperature-density ($T - \rho$) diagram of two-dimensional (2D) and three-dimensional (3D) L-J system for the dispersity index of 5%.

The plot shows that with increasing temperature the density of solid gradually decreases and solid goes to liquid phase at T* = 1.12 for 2D system. In addition, the plateau joining high density solid branch and low density liquid branch represents the solid-liquid coexistence line at which solid and liquid phases coexist during the melting transition. For 3D system, the solid goes to liquid phase at T* = 1.40 and solid-liquid coexistence line is also shown in the plot.

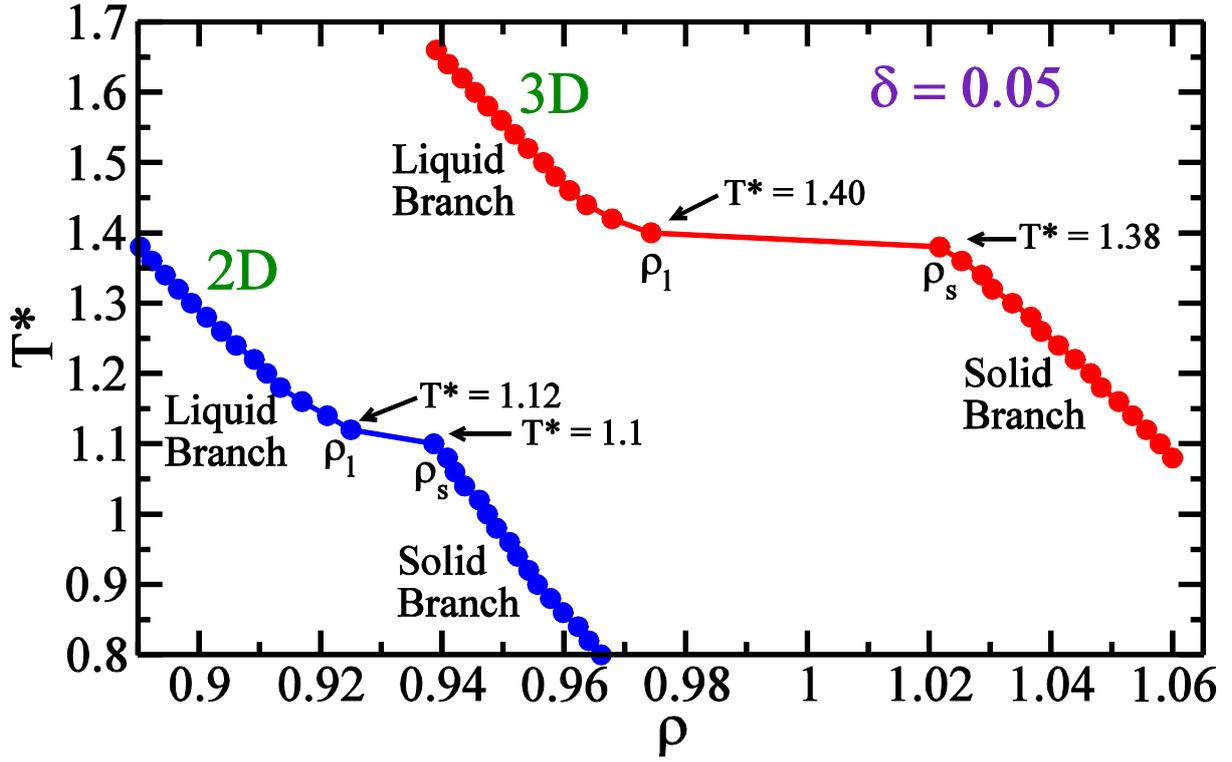

Figure 1: The temperature (T) versus density ($\rho$) plot in the case of two-dimensional (2D) and three-dimensional (3D) L-J system for polydispersity index $\delta = 0.05$. The high density solid branch and low density liquid branch is joined by a plateau with two end points $T^* = 1.1$ and $T^* = 1.12$ respectively for 2D system. On the other hand, the high density solid branch and low density liquid branch is joined by a plateau with two end points $T^* = 1.38$ and $T^* = 1.40$ respectively for 3D system.

In order to understand the detail picture of solid-liquid melting transition in 2D and 3D polydisperse systems, systematic studies have been carried out and the results are presented in the subsequent section.

## 2.1.2. Bond order parameter: Quantification of solid-liquid melting transition in 2D and 3D polydisperse systems

We present here the change of order parameters with respect to increase in temperature during temperature induced melting transition both for 2D and 3D systems in order to quantify the melting phenomenon. For two-dimensional system, global bond orientation parameter is defined as,

$$\Psi_6^g = \frac{1}{N} \left| \sum_{k=1}^{N} \frac{1}{n_k} \sum_{j=1}^{n_k} e^{i6\theta_{kj}} \right|$$

where N is the number of particles in the two-dimensional layer and $n_k$ denotes number of $K$ nearest neighbors. $\theta_{kj}$ is the angle between central particle k and nearest neighbor $j$. Two particles are considered as nearest neighbours if $r_{ij} < r_{min}\sigma_{ij}$ for their center-to-center separation. The value of $\Psi_6^g$ is equal to 1 for a perfect hexagonal plane, but is far from 1 for a disordered phase. A significantly large value of order parameter indicates that there is long range translational order and the structure is solid like and similarly a low value of order parameter indicates that the structure is disordered and liquid like.

For 3D system, the bond order is introduced by Steinhardt *et al.* [33]. Following their definition, a vector $r_{ij}$ pointing from a given molecule (i) to one of its nearest neighbor (j) is denoted as a "bond". For each bond one determines the quantity as,

$$Q_{lm}(r_{ij}) = Y_{lm}\left[\theta(r_{ij})\phi(r_{ij})\right]$$

where, $Y_{lm}[\theta(r_{ij})\phi(r_{ij})]$ is spherical harmonics. $\theta(r_{ij})$ and $\phi(r_{ij})$ are the polar and azimuthal angles of vector $r_{ij}$ respectively with respect to an arbitrary reference frame. The global bond order is obtained by averaging over all bonds in the system,

$$\bar{Q}_{lm} = \frac{1}{N_b}\Sigma_{bonds} Q_{lm}(r_{ij})$$

where, $N_b$ is the number of bonds. To make the order parameters invariant with respect to rotations of the reference frame, the second-order invariants are defined as,

$$Q_l = \left[\frac{4\pi}{2l+1}\sum_{m=-l}^{m=l}\bar{Q}_{lm}^2\right]^{1/2}$$

In FCC lattice the orientational order is represented in terms of six-fold symmetry corresponding to $l = 6$ and for perfect FCC crystal, $Q_6 = 0.5745$. We have calculated global bond order parameters with increasing temperature both for 2D and 3D systems. **Fig. 2** represents the change in $\Psi_6^g$ and $Q_6$ with increasing temperature for 2D and 3D, respectively. It is evident from the result that with increasing temperature T*, the values of $\Psi_6^g$ and $Q_6$ slightly decrease. Interestingly, $\Psi_6^g$ shows a sharp structural transition at T* = 1.1 for 2D system and $Q_6$ shows a sharp structural transition at T* = 1.38 for 3D. These sharp changes in $\Psi_6^g$ and $Q_6$ values, in 2D and 3D systems respectively, are the clear evidences of solid to liquid transition.

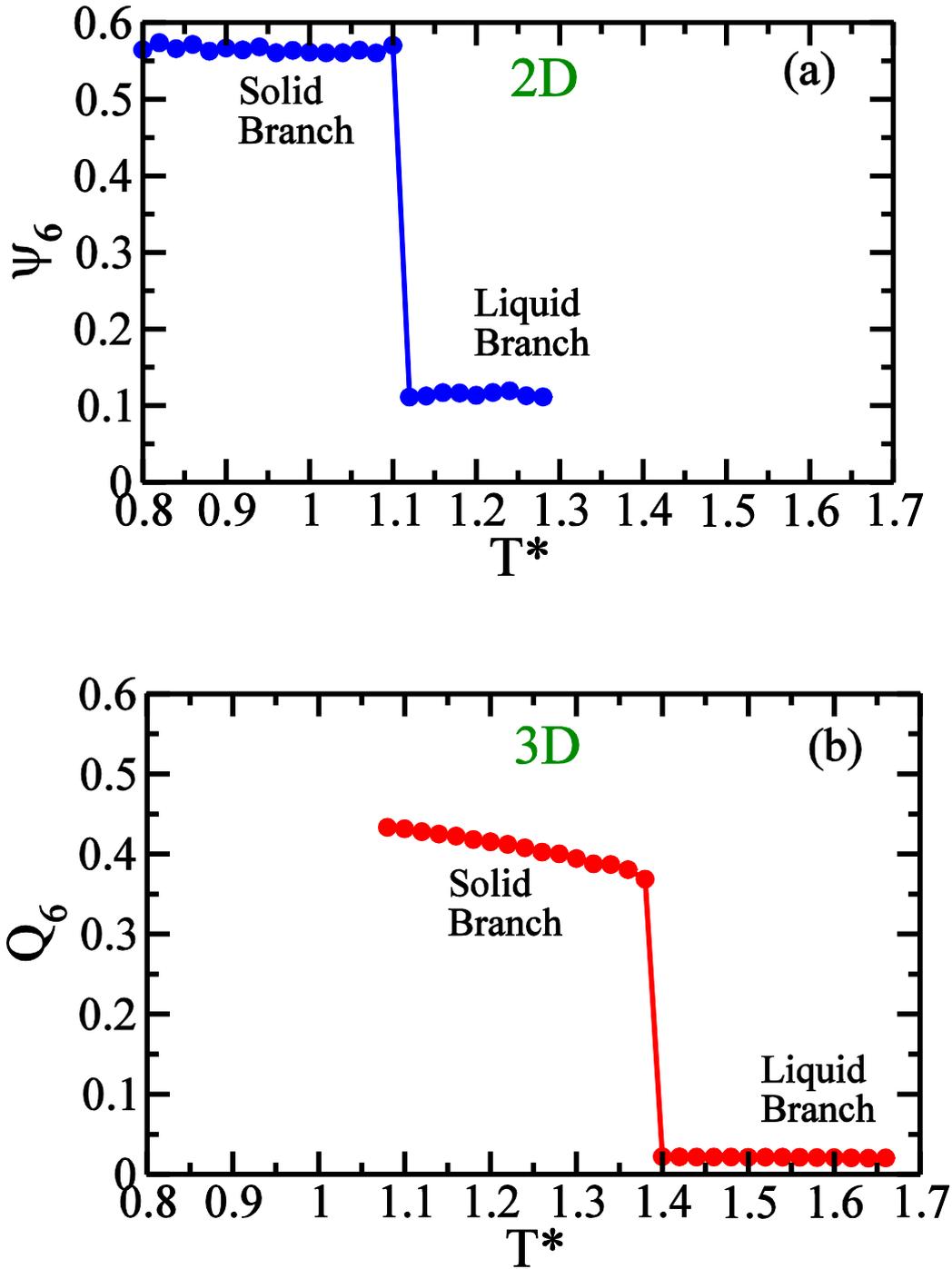

Figure 2: (a) Change of order parameters versus temperature during temperature induced melting transition in 2D polydisperse system. Please note that for two dimensional system, $\Psi_6^g$ shows a sharp structural transition at T* = 1.1 signifying the melting transition. (b) Change of order parameters versus temperature during temperature induced melting transition in 3D

system. Please note that for three dimensional system, $Q_6$ shows a sharp structural transition at T* = 1.38 signaling the solid-liquid transition.

## 2.2. Size dependent Lindemann constant

The Lindemann parameter can be defined as,

$$L = \frac{\sqrt{\langle \Delta r^2 \rangle}}{a}$$

where, $L$ is the Lindemann parameter for the associated polydisperse system and $a$ is the mean distance between particles and $\Delta r = |r_i(t) - R_i|$; $r_i(t)$ is the instantaneous position of atom $i$ and $R_i$ is the equilibrium position of atom $i$. The mean distance between the particles can be considered as the position of the first peak of the radial distribution function. The threshold value of Lindemann parameter for melting of bulk face-centered cubic (FCC) crystal is 0.22 and for body-centered cubic (BCC) crystal is 0.18 [34-37]. Chakravarty *et al* demonstrated Lindemann measures for the solid-liquid phase transition based on the positional displacements of atoms from their locations in the corresponding mechanically stable inherent structures in the neighborhood of the melting transition for a Lennard-Jones-type solid [1].

In our study, in order to understand the role of small and big sized particles near melting temperature, we have computed the Lindemann parameter of small and big sizes particles with increasing temperature along the melting line for both 2D and 3D polydisperse systems. It is to be noted that the diameter range of the particles has been divided into two types: the particles having diameter range less than one diameter are defined as small particles and particles having diameter range greater than or equal to one diameter are defined as big particles. The computed Lindemann parameters of small and big sized particles with increasing temperature are presented in **Fig. 3 (a)** and **Fig. 3 (b)** for 2D and 3D polydisperse system respectively. Near melting temperature, the values of Lindemann parameters for small particles are found to be higher as compared to big particles. This implies that small particles take part in melting at first and later big particles undergo slow melting.

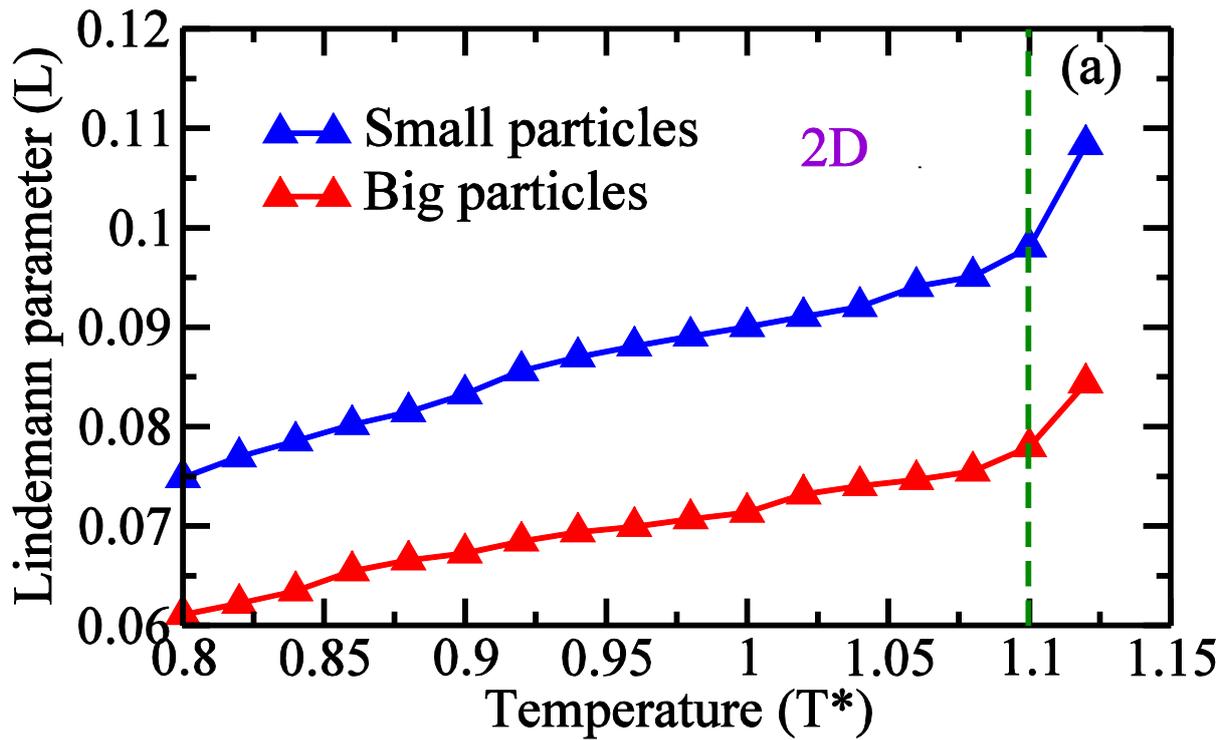

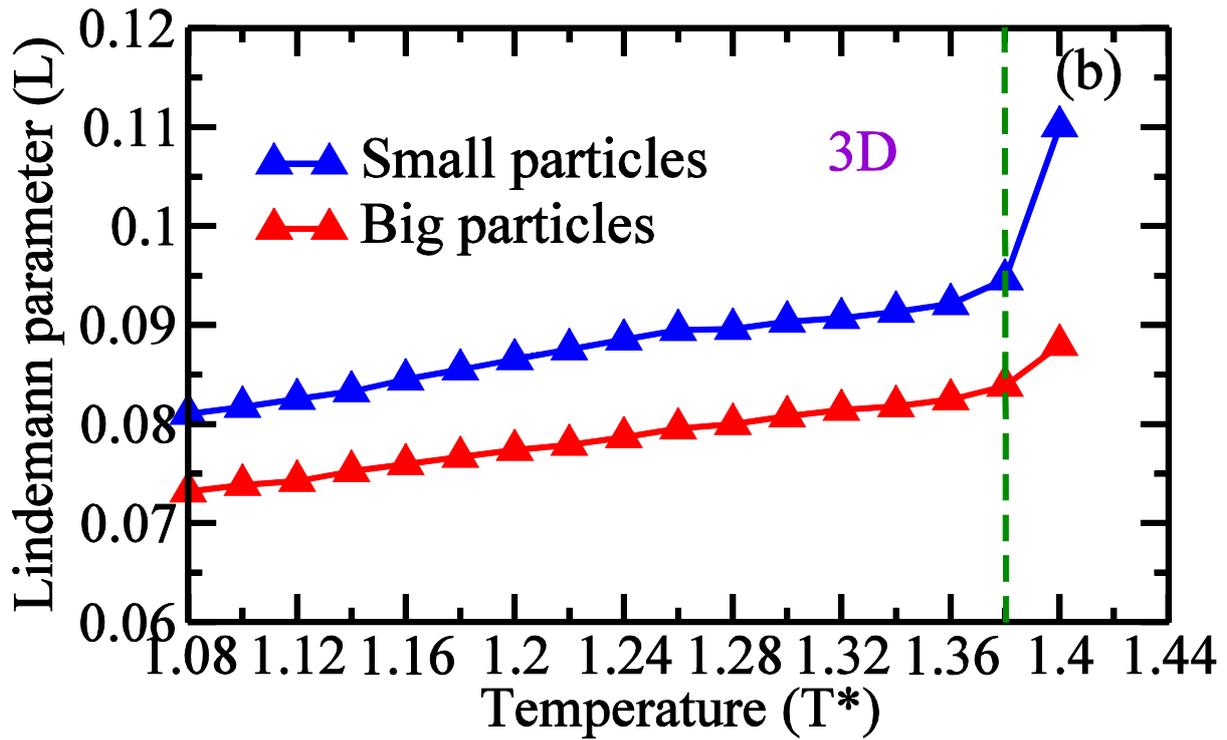

Figure 3: Lindemann parameter versus temperature plot for small and big particles having polydispersity index $\delta = 0.05$ during melting in case of (a) 2D and (b) 3D system. The values of Lindemann parameters for small particles are higher compared to big particles near melting temperature and subsequently the values of Lindemann parameter reach the threshold value for melting. This signifies that small particles take part during the melting process in the beginning and afterwards big particles experience slow melting. Please note that these figures [Fig. 3(a) and (b)] may be compared with Fig. 2 from the work by Chakravarty *et al.* [1].

## 2.3. Behavior of small and big particles before melting: Segregation followed by melting

Segregation is a very well-known and common phenomenon that occurs before melting/freezing transition where segregation is caused by density differences in the melting components. This density difference occurs due to different diffusion rate of melt components. Even at near melting temperature, small particles show the high diffusion than big particles. This raises a question about the behavior of small and big particles before melting. Although several studies have already been reported on segregation for various other systems but for size induced polydisperse system the behavior of the particles before melting still is not clearly understood as it is a complex system. In our study, in order to understand the behavior of particles before melting, a detailed trajectory analysis has been carried out at near melting temperature (T* = 1.1 for 2D, T* = 1.36 for 3D) and our results reveal that segregation occurs both in 2D and 3Dbefore melting. The trajectories before and after melting are presented by taking snapshots as shown in **Fig. 4(a)**, **Fig. 4(b)** for 2D and **Fig. 5(a)**, **Fig. 5(b)** for 3D respectively. It is evident from the **figures 4(a)**, **4(b)**, **5(a)** and **5(b)** that the segregation occurs before melting both in 2D and 3D. However, we observed that the degree of tendency for the segregation of the small particles is higher than the big particles.

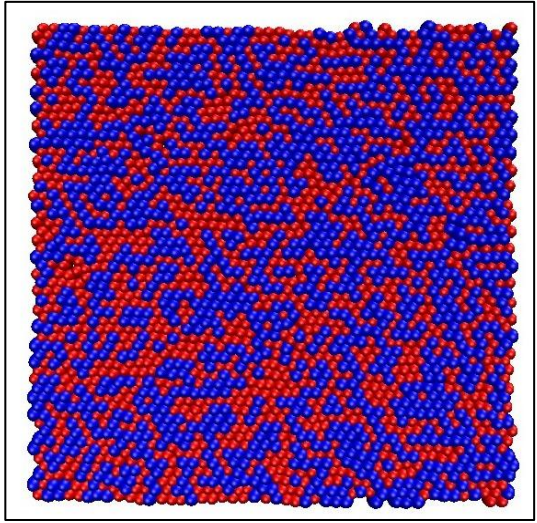 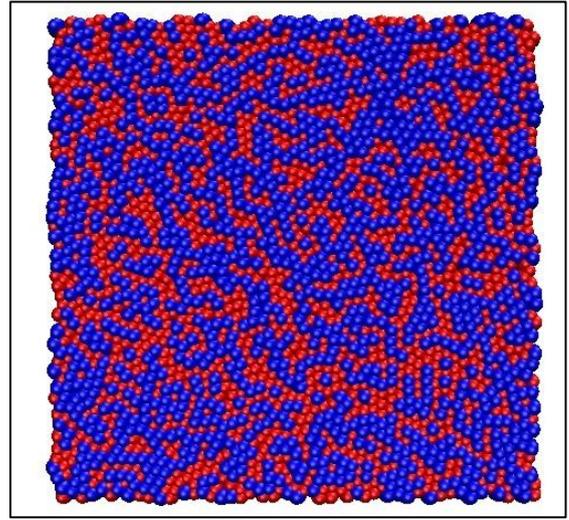

(a) (b)

**Figure 4: Snapshots for 2D showing the spatial positions of small particle in red and big particle in blue color. (a) Please note that the structure is taken before melting at T\* = 1.1 and (b) the structure is taken after melting at T\* = 1.12. The snapshot before melting clearly shows that the segregation occurs before melting.**

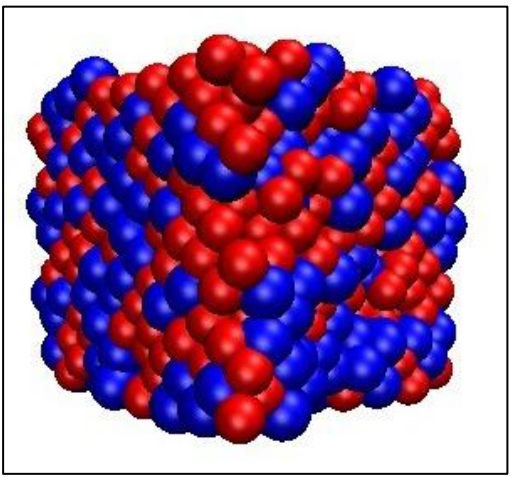 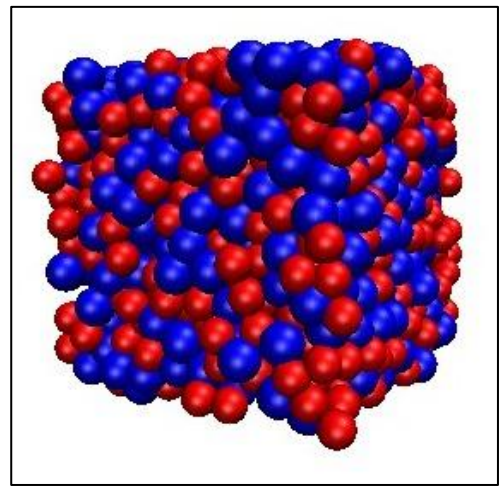

(a) (b)

**Figure 5:** Snapshots for 3D showing the spatial positions of small particle in red and big particle in blue color. (a) Please note that the structure is taken before melting at T* = 1.38 and (b) the structure in is taken after melting at T* = 1.40. The snapshot before melting clearly shows that the segregation occurs before melting.

In order to characterize the segregation, we have calculated the change in the distance between small particles and between big particles with respect to time at T* = 1.1 for 2D as shown in **Fig. 6(a)** and at T* = 1.36 for 3D as shown in **Fig. 6(b)**. The small-small and big-big particle distances decrease sharply with time as shown in **Fig. 6(a)** and **Fig. 6(b)**. This indicates that the segregation between small particles as well as between big particles occurs before melting. However, the tendency of segregation between small-small particles is higher than that of big-big particles.

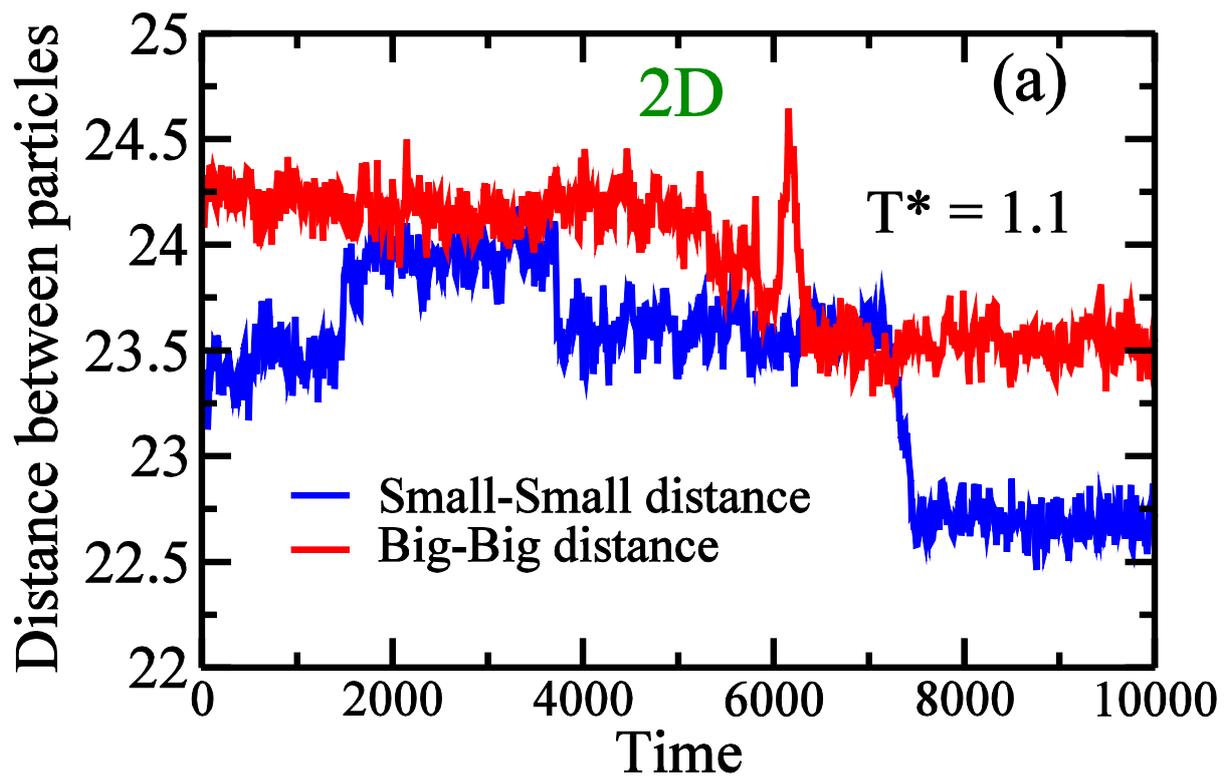

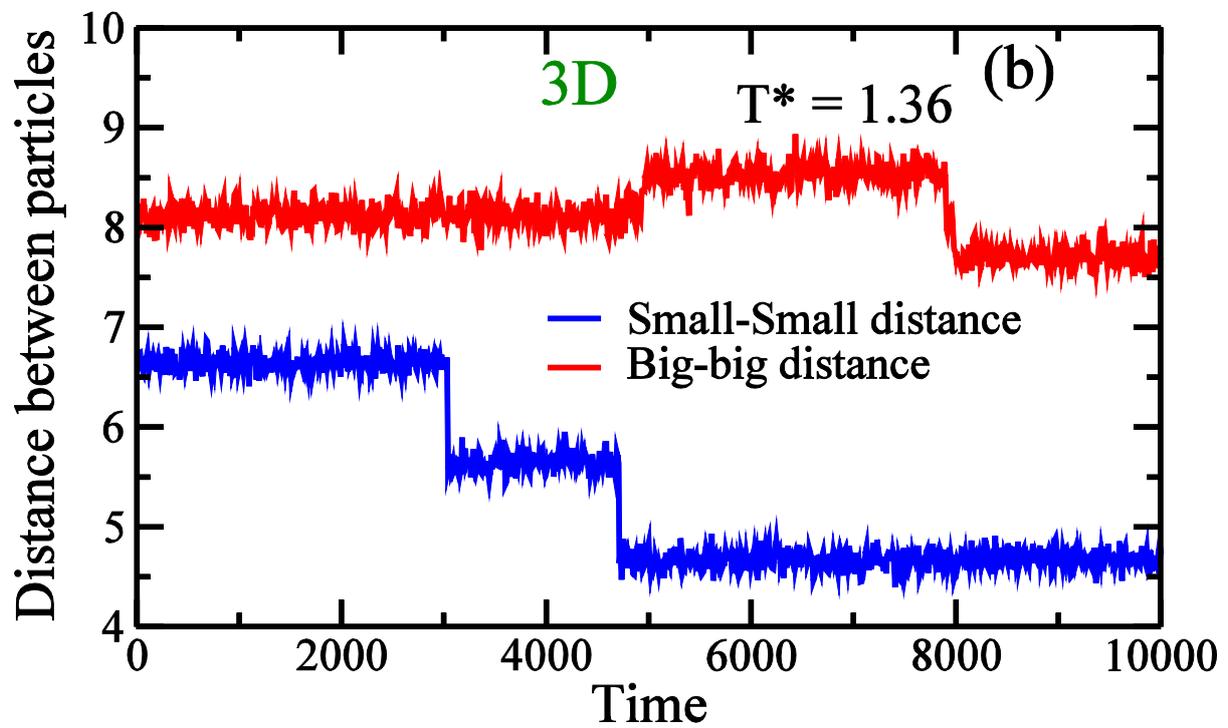

**Figure 6:** Evolution of distance between small particles and between big particles with respect to time for (a) two-dimensional (2D) and (b) three-dimensional (3D) polydisperse systems. Distance between small-small particles is presented in blue and that of big-big particles is presented in red colour. The distances presented here are at melting temperature $T^* = 1.1$ for 2D and $T^* = 1.36$ for 3D polydisperse systems.

## 3. Conclusion

We have calculated the size dependent mean square displacement as well as Lindemann parameter for small and big sized particles both in the cases of two dimensional and three dimensional polydisperse systems. Our results reveal that as we approach the solid-liquid melting line from lower temperature at constant pressure, an increase of temperature gives rise to a significantly larger increase in the mean square displacements of smaller sized particles than that of the bigger sized particles. In fact, this effect manifests itself most clearly near the melting temperature. One can surmise that near the melting temperature the small size particles show higher tendency of melting compared to that of the bigger particles, and therefore display sudden jump in the Lindemann ratio before the thermodynamic melting temperature. This implies that the polydisperse system shows bimodal distribution at near melting temperature. This is accompanied by segregation between small particles or between big particles to occur before melting.

Although the segregation phenomenon is common before melting of polydisperse solids, a quantification of this segregation is rather difficult for the complex polydisperse system. We have been able to demonstrate that near melting of two dimensional solid, the relative distance between small-small and big-big particles decreases while in the 3-D solid, that between small and small particles decreases, that between big and big does not show any noticeable trend. We have not been able to find any convincing reason of this difference yet. There are several other notable features observed in the present study.

(1) The Lindemann ratio (L) obtained from root mean square displacement (RMSD) is rather strongly size dependent signaling the breakdown of universal Lindemann criterion of melting.

(2) Both in 2D and 3D, the larger diffusion of small particles do not lead to lessening of the sharpness of melting. They both melt with the lattice.

(3) The observed larger value of diffusion of smaller particle is in apparent agreement with Stokes-Einstein relation.

(4) Segregation is found to happen first and subsequently followed by melting.

We shall explore in a later work that the results of the existing studies (both of ours and of Chakravarty et al.'s) can be explained using Fixman's self-consistent phonon theory [38-40].

**Acknowledgement:** We would like to thank Prof. Stuart A. Rice from the University of Chicago for scientific discussions. This work was supported in parts by grants from DST, BRNS and CSIR (India). B.B. thanks DST for support through J.C. Bose Fellowship.